\documentclass[aps,amssymb,prl, amsmath, twocolumn, floatfix]{revtex4}
\usepackage{graphics}
\usepackage{psfrag}
\usepackage{epsfig}
\usepackage{subfigure}

\begin{document}
\bibliographystyle{prsty}

\title{Protein folding on rugged energy landscapes: Conformational diffusion on fractal networks}
\author{Gregg Lois}
\author{Jerzy Blawzdziewicz}
\author{Corey S. O'Hern}
\affiliation{
Department of Mechanical Engineering and Department of Physics, 
Yale University, New Haven, Connecticut 06520-8286
}

\begin{abstract}
We employ simulations of model proteins to study folding on
rugged energy landscapes. We construct ``first-passage''
networks as the system transitions from unfolded to native states.
The nodes and bonds in these networks correspond to basins and
transitions between them in the energy landscape. We find
power-laws between the folding time and number of nodes and
bonds. We show that these scalings are determined by the fractal
properties of first-passage networks.  Reliable folding is possible in
systems with rugged energy landscapes because first passage networks
have small fractal dimension.
\end{abstract}
\maketitle

Understanding how proteins reliably fold to their native conformations
despite frustration in the form of non-native interactions between
residues is an important, open question.  Advances in experimental
techniques, such as single-molecule fluorescence~\cite{eatonscience}
and fast thermal quenching methods~\cite{fastfolding}, have enabled a
quantitative characterization of the dynamics that occur during
folding of single proteins.  For example, we now know that a large
number of metastable conformations are sampled during the folding and
unfolding processes, as observed in folding stability~\cite{tertiary}
and mechanical denaturation~\cite{pulling} studies.

\begin{figure}[t]
\begin{center}
\scalebox{0.32}{\includegraphics{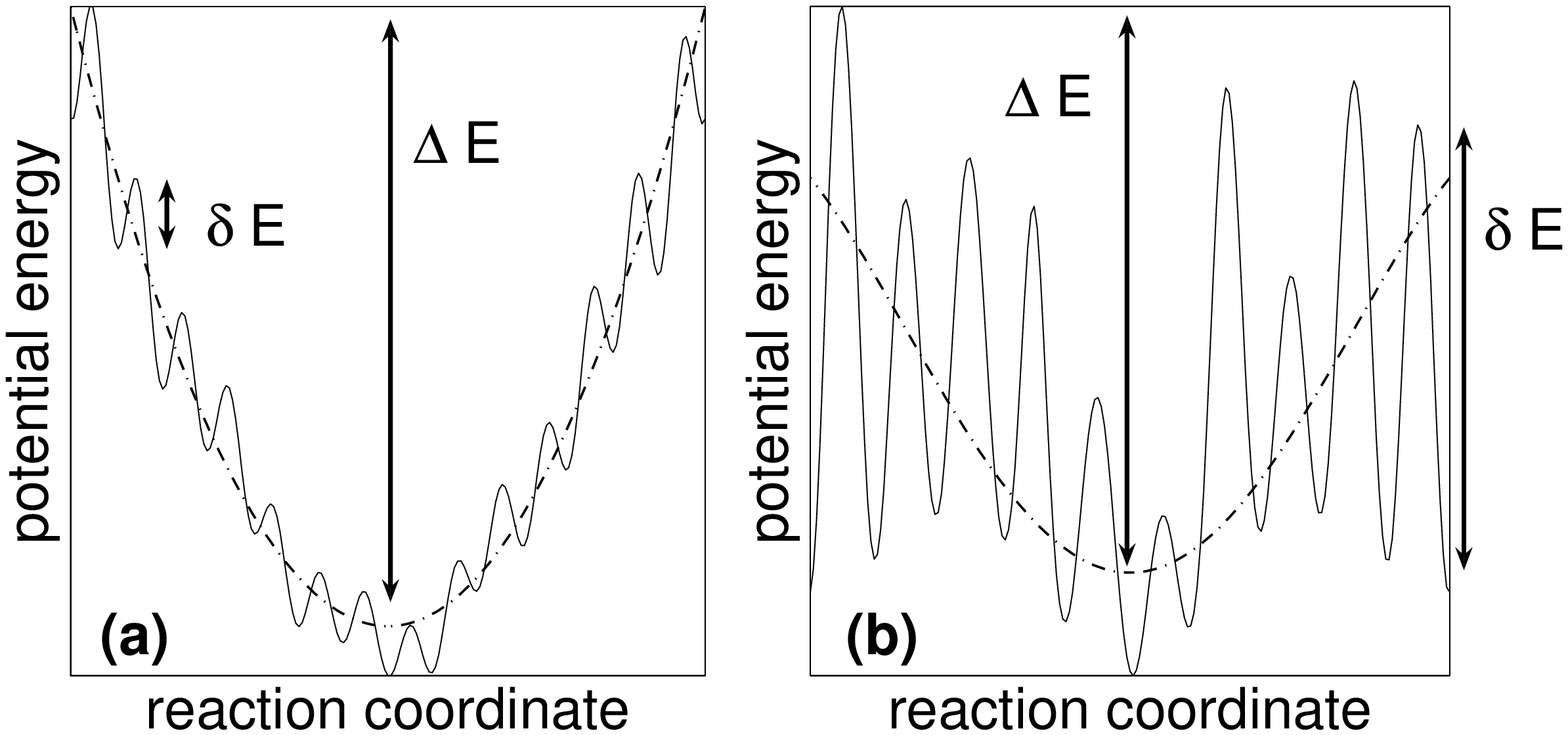}}
\vspace{-0.15 in}
\caption{ \label{diagram} Schematics of (a) funneled and (b)
rugged energy landscapes.  In (a), the depth of the energy 
minimum that drives folding $\Delta E \gg \delta E$, where $\delta 
E$ gives the root-mean-square energy fluctuations over the given range of the 
reaction coordinate. In (b), $\Delta E \sim \delta E$.}
\end{center}
\vspace{-0.3 in}
\end{figure}

How does a protein fold reliably to its native conformation even
though a large number of metastable states exist?  For over twenty
years the answer to this question has been the principle of minimal
frustration~\cite{minimalfrustration}.  Within this framework, one
recognizes that metastable states are present, but assumes that the
barriers separating local energy minima are
sufficiently low that there is still a large thermodynamic force
driving folding to the native state~\cite{funnel}.  This idea is
illustrated by the funneled energy landscape in
Fig.~\ref{diagram} (a), where the roughness scale $\delta E$ is much
smaller than depth of the energy minimum $\Delta E$ that drives
folding ($\delta E \ll \Delta E$).  While the funneled energy
landscape may explain how some proteins fold
reliably~\cite{bryngelson}, a different picture, {\it i.e.} rugged
energy landscapes may describe folding in metastable
~\cite{metastablefold} and intrinsically
disordered~\cite{pappu} proteins, as well as misfolding~\cite{csoto}.
Rugged energy landscapes, as shown in Fig.~\ref{diagram} (b), possess
a roughness scale that is comparable to that of the smooth funnel
$\delta E \sim \Delta E$.  In this limit, the thermodynamic drive to
fold is absent on biological timescales, and protein conformational
dynamics proceed via activation over energy barriers with
only local knowledge of the landscape.

What physical observables differentiate proteins with funneled versus
rugged landscapes?  This is a difficult question to answer since,
although funneled energy landscapes have been studied extensively,
virtually no research has focused on reliable folding in proteins with
rugged energy landscapes.  We make a crucial first step in answering
this question by studying the properties of a model protein that
reliably folds to its native state on a rugged energy landscape with
$10^2-10^4$ distinct basins sampled during folding.  (A basin is a
region of configuration space, or collection of conformations, that
relaxes to a single local energy minimum when thermal fluctuations are
suppressed~\cite{stillinger}.)  Instead of discrete pathways through
the energy landscape, we find a statistical ensemble of pathways with
large fluctuations in folding times.  The folding time and number of
distinct basins sampled during folding scale as a power-law, which
suggests that reliable folding on rugged landscapes can be described
as conformational diffusion on a {\em fractal} network of basins.

{\bf Heteropolymer model: } To study proteins with rugged energy
landscapes, simulation models should possess three key features: (1)
unique native state, (2) many metastable, local energy minima, and (3)
large energy barriers that separate local minima so that $\delta E
\sim \Delta E$.  Further, we must be able to search configuration
space in a reasonable amount of computer time, which excludes all-atom
simulations. In these studies, we will focus on a model heteropolymer
that exhibits features (1)-(3).

\begin{figure}[t]
\begin{center}
\scalebox{0.67}{\includegraphics{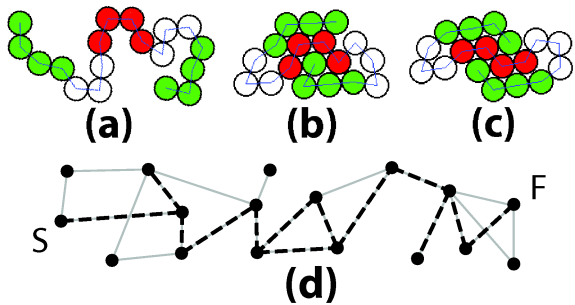}}
\vspace{-0.15 in}
\caption{ \label{pictures} (Color online). The heteropolymer model
in its (a) extended, (b) metastable misfolded, and (c)
native states.  (d) Schematic of a first-passage network (black dashed
lines) from basin `S' to `F', superimposed on the complete network
composed of all basins and transitions between them (gray lines).}
\end{center}
\vspace{-0.3 in}
\end{figure}

We model proteins as heteropolymers composed of equal-sized spherical
monomers with hydrophobic and hydrophilic
interactions~\cite{thirumalai}.  The model includes hydrophilic
monomers (white) and two types of hydrophobic monomers (red and green)
as shown in Fig.~\ref{pictures}.  Green and red monomers interact via
an attractive Lennard-Jones potential with minimum energy
$-E_\mathrm{att}$, except the green monomers on both ends of the chain
that interact with minimum energy $-2 E_\mathrm{att}$.  All other
monomer-monomer interactions are purely repulsive~\cite{ferp}.  We
also include a FENE potential~\cite{fenepotential} between adjacent
monomers to maintain the polymer constraint.  We simulate the
$18\mathrm{-mer}$ sequence ggggwwwrrrrwwwgggg, where g,w and r
represent green, white and red monomers, respectively.  This model
displays a complex energy landscape with $\sim 10^5$ distinct local
energy minima. For simplicity, local minima are defined by the
list of contacting green and red monomers~\cite{footnote2}.  The
native conformation of this heteropolymer is given by the
particular set of $14$ green-red contacts shown in
Fig.~\ref{pictures} (c).

\begin{figure}[t]
\begin{center}
\scalebox{0.32}{\includegraphics{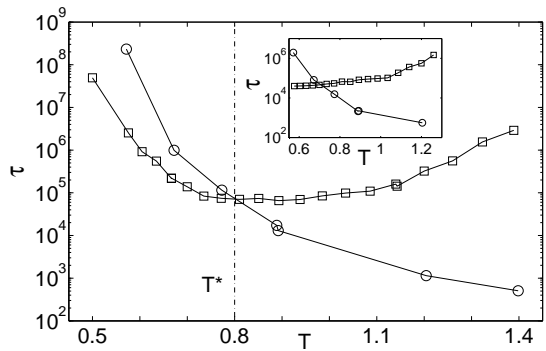}}
\vspace{-0.15 in}
\caption{ \label{chevron} Ensemble-averaged folding $\tau_f$ (squares)
and unfolding $\tau_u$ (circles) times vs temperature for the
heteropolymer (main figure) and Go (inset) models.  The vertical line
at $T^*=0.8$ indicates the folding temperature.}
\end{center}
\vspace{-0.3 in}
\end{figure}

Thermal fluctuations of the heteropolymer are studied using Brownian
dynamics, where the temperature $T$ is reported in units of the
attractive energy, {\emph e.g.}  $T=1/3$ corresponds to thermal energy
$E_\mathrm{att}/3$.  To compare results for rugged and funneled energy
landscapes, we also simulated the same heteropolymer with
Go-interactions~\cite{gomodel}, where attractive interactions are only
included between monomers that form contacts in the native state.  The
simplest measures of kinetics are the folding and unfolding times
shown in Fig.~\ref{chevron}.  The folding time $\tau_f$ is calculated
by preparing the heteropolymer in an ensemble of extended states and
measuring the average folding time to the native state.  $\tau_u$ is
the average unfolding time from the native state to any extended state
with zero red-green contacts.  For temperature $T<T^*= 0.8$, $\tau_f <
\tau_u$, and the extended conformation is significantly less stable
than the native state.  The increase in $\tau_f$ as $T$ decreases, as
shown in Fig.~\ref{chevron}, has been observed in experimental studies
of proteins~\cite{segawa} and is a general feature of materials
quenched below the glass transition~\cite{upglass} when energy
barriers become large compared to $T$.  An important feature of the
heteropolymer model is that folding only occurs for temperatures where
$d \tau_f/dT <0$.  In contrast, folding simulations of the Go-model
yield $d \tau_f/dT>0$ for all $T$, as shown in the inset to
Fig.~\ref{chevron}.

{\bf First-passage networks: } For each heteropolymer conformation, we
can determine the list of contacting green and red monomers and
uniquely associate this list of contacts with a basin that surrounds
the associated local energy minimum.  For rugged landscapes, the
system will sample a large number of basins as folding proceeds from
the extended to the native state.  The trajectory of the model protein
as it folds can be viewed as a network of connected nodes in
configuration space.  The nodes represent the basin of a local energy
minimum sampled by the system, and bonds that join two nodes represent
transitions from one basin to another.  These networks are termed
``first-passage networks'' since they are formed as the protein makes
its first passage from an initial to the native conformation.  Note
that each first-passage network is a subset of all basins and
transitions between them, as illustrated in Fig.~\ref{pictures} (d).

We compiled $\sim 10^6$ first-passage networks originating from
the non-native conformation in Fig.~\ref{pictures} (b) and ending at the
native state over a range of $T\leq 0.8$.  We map the conformation of
the heteropolymer to its associated basin every $q$ time steps to
construct first-passage networks.  We assume that the features of
the first-passage networks depend on $T$ but are independent of the
initial state since the first-passage networks are composed of a large number 
of nodes.

\begin{figure}
\begin{center}
\scalebox{0.32}{\includegraphics{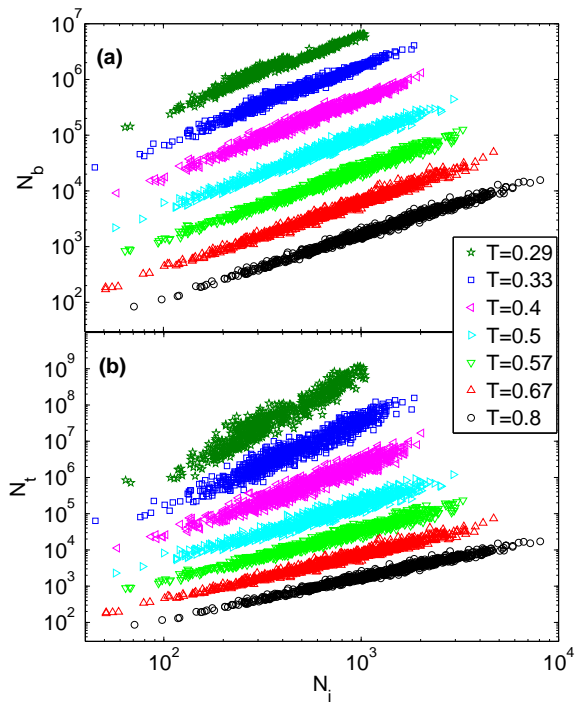}}
\vspace{-0.15 in}
\caption{ \label{scalings} (Color online). Number of (a) bonds $N_b$
and (b) transitions $N_t$ in first-passage networks vs the number of
nodes $N_i$ over a range of temperature.  For each $T$, $N_b$ and
$N_t$ have been multiplied by constant factors (shifted vertically)
for clarity.
}
\end{center}
\vspace{-0.3 in}
\end{figure}

The simplest properties of first-passage networks are the number of
distinct basins sampled (nodes) $N_i$ and bonds $N_b$.  Nodes and
bonds are only counted once, even if multiple transitions are made
between a given set of basins.  We also measure the total number of
transitions $N_t \propto \tau_f \geq N_b$.  Fig.~\ref{scalings} shows
raw data for the number of bonds $N_b$ and transitions $N_t$ plotted
versus the number of nodes $N_i$ using $q=1000$.  There are
$850$ data points for each temperature, each taken from a distinct
first-passage network. For all $T$ the number of sampled basins,
$N_i$, fluctuates between $10^2$ and $10^4$, which indicates that the
model protein adopts a large number of conformations before arriving
at the native state.  The wide range of $N_i$ indicates that there is
not a single folding pathway, but rather a statistical ensemble of
pathways.

In Fig.~\ref{scalings}, $N_b$, $N_t$ and $N_i$ show strong
fluctuations from one realization to the next; however, the fluctuations obey
power-law scaling: 
\vspace{-0.05 in}
\begin{equation}
N_b \propto N_i^{\, \,\Lambda}~~~{\rm and}~~~N_t \propto N_i^{\, \, \Gamma }. 
\label{defgamma}
\vspace{-0.05 in}
\end{equation}
This correlation is non-trivial and depends on global properties of
first-passage networks.  We find that distributions of local features 
of the network, such as single-jump activation times and distances, and 
the number of bonds per node, are exponential.
Thus, local properties of first-passage networks cannot be responsible
for the power-law scaling.

\begin{figure}
\begin{center}
\psfrag{oover}{\Huge{$1/\kappa \, d_\mathrm{f}$}}
\scalebox{0.32}{\includegraphics{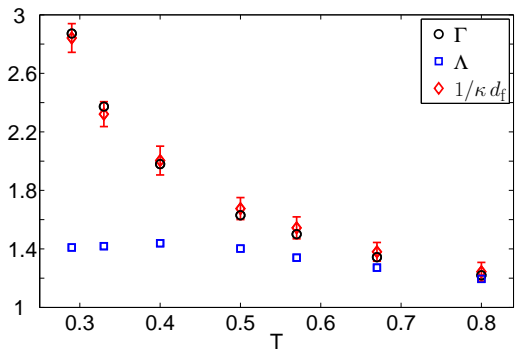}}
\vspace{-0.15 in}
\caption{ \label{gammalambda} (Color online). The scaling exponents
$\Gamma$ and $\Lambda$ and the prediction $1/\kappa
d_\mathrm{f}$ for $\Gamma$ from Eq.~\ref{eqngamma}. Error bars for
$\Gamma$ and $\Lambda$ are smaller than the symbol size.  }
\end{center}
\vspace{-0.3 in}
\end{figure}

In Fig.~\ref{gammalambda}, we plot the scaling exponents $\Gamma$ and
$\Lambda$ at different temperatures $T$.  While $\Lambda$ reaches a
plateau at $\approx 1.4$ at small $T$, $\Gamma$ continues to increase
with decreasing $T$. The increase of $\Gamma$ is a signature of
temperature-dependent exploration of configuration space in systems
with rugged landscapes.  A system with a rugged energy landscape at
energy $E$ only samples a small temperature-dependent fraction of
conformations at that energy due to large activation barriers.  In
contrast, $\Gamma \approx 1.5$ at all $T$ for the same heteropolymer
model with Go-interactions.  In systems with funneled energy
landscapes ({\it i.e.} the Go model), a protein with energy $E$
samples conformations with that energy more uniformly.

The data shown in Fig.~\ref{scalings} are obtained by identifying
basins every $q=1000$ time steps.  We have also performed
simulations in the range $1<q<10^4$ and observe that the
exponents $\Gamma$ and $\Lambda$ are independent of $q$.  These
results further indicate that first-passage networks are self-similar and 
fractal.

{\bf Origin of power laws:} If we assume that first-passage networks
are fractal, we can predict the exponent $\Gamma$ from the fractal
scaling exponents of the network.  This assumption will be verified
{\it a posteriori}.  

On any network we can define the chemical
distance $\Delta c$ given by the shortest path between two nodes of
the network.  This distance is useful because it depends only on
network connectivity and is independent of the embedding
space~\cite{footnote1}.  For a fractal network, we
expect~\cite{dfractal}
\begin{eqnarray}
\label{fractaldyn}
\Delta c &\propto& t^{\,\kappa}, \\
N(\Delta c) &\propto& \Delta c^{\,d_\mathrm{f}},
\label{fractaldim}
\end{eqnarray}
where $N(\Delta c)$ is the number of distinct basins sampled within
chemical distance $\Delta c$ and time interval $t$, $d_f$ is the chemical
fractal dimension, and the exponent $\kappa$ characterizes the scaling of 
chemical distance with time.

\begin{figure}
\begin{center}
\psfrag{kp}{\Huge{$\kappa$}}
\scalebox{0.32}{\includegraphics{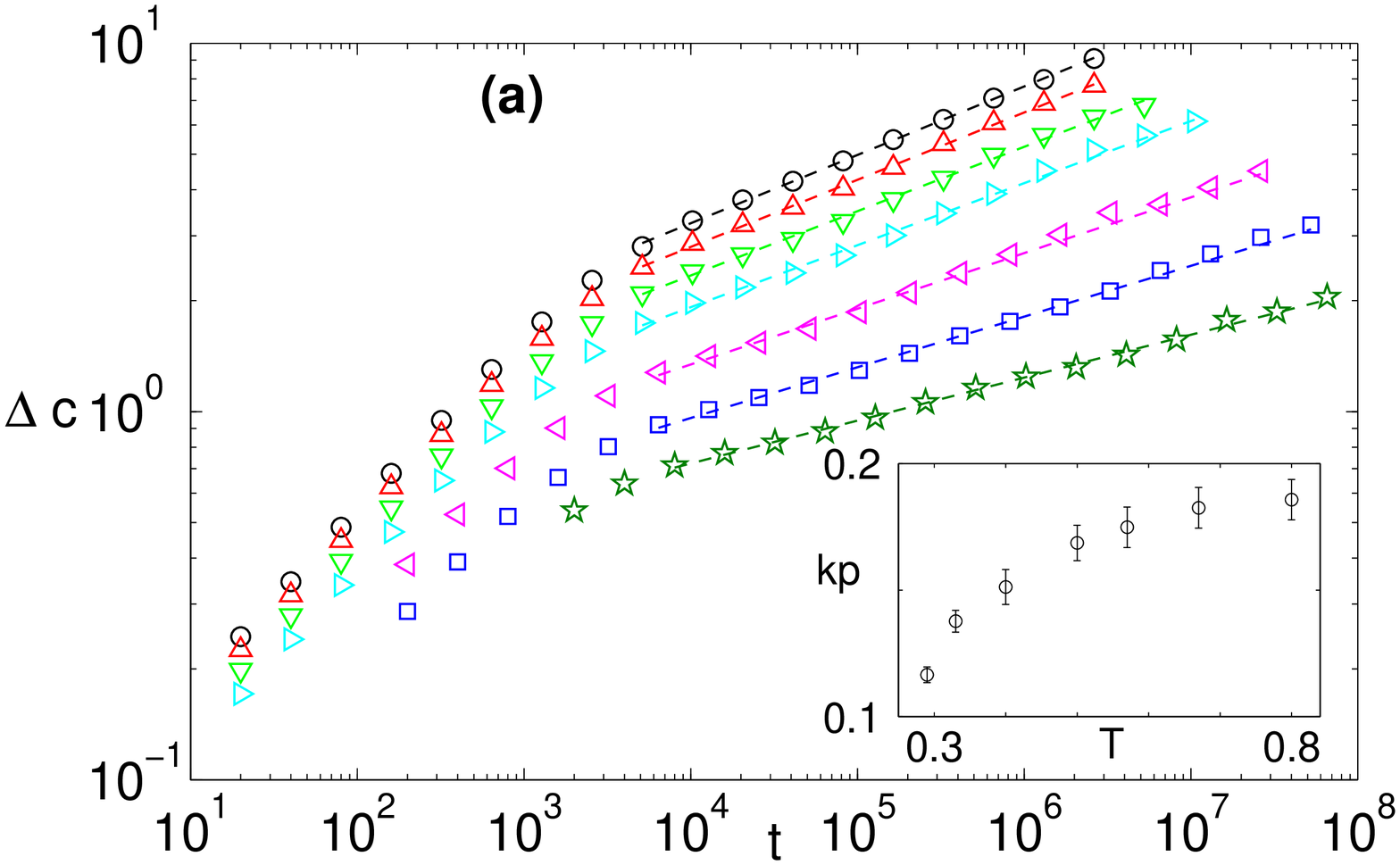}} 
\scalebox{0.32}{\includegraphics{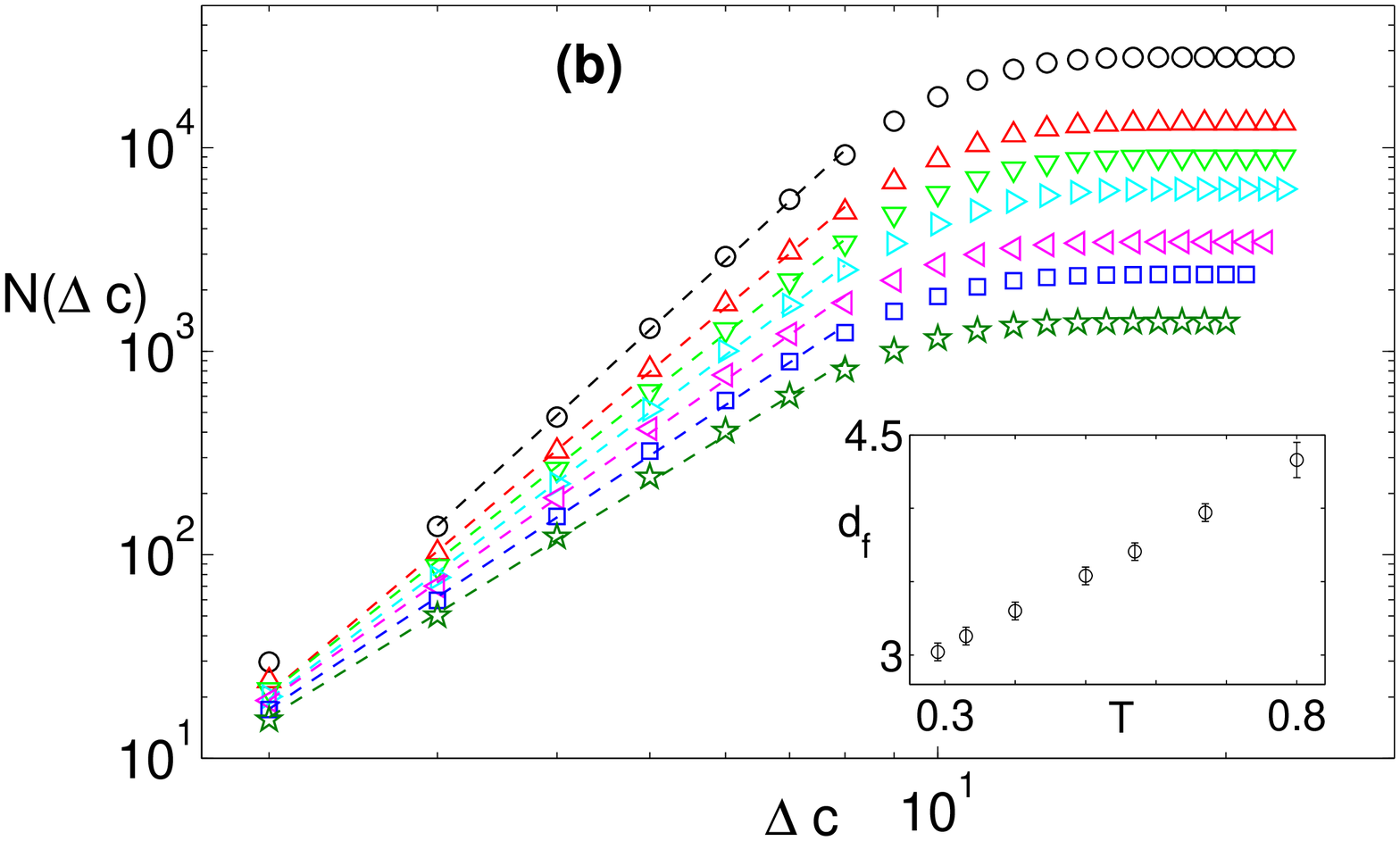}} 
\vspace{-0.15 in}
\caption{ \label{fractal} (Color online). (a) The mean chemical
distance $\Delta c$ sampled in the time interval $t$ by the
heteropolymer and (b) the mean number of basins $N(\Delta c)$ within
$\Delta c$ at different temperatures.  In (a) and (b), the symbols are
the same as in Fig.~\ref{scalings}, and the insets display the scaling
exponents used to fit the data (dotted lines) for different
temperatures.  }
\end{center}
\vspace{-0.3 in}
\end{figure}

Given these relations, the correlation between $N_i$ and $N_t$ can be
explained as follows.  A single first-passage network is formed over
folding time $\tau_f \propto N_t$, during which the system explores
average chemical distance $\Delta c \propto N_t^{\,\,\kappa}$
(Eq.~\ref{fractaldyn}).  Moreover, for a given chemical distance
$\Delta c$, the number of sampled basins on the first passage network
scales as $N_i \propto N(\Delta c) \propto \Delta c^{d_\mathrm{f}}$
(Eq.~\ref{fractaldim}).  Thus, both $N_i$ and $N_t$ are related to
$\Delta c$, and we find $N_t \propto N_i^{\, \, 1/\kappa
d_\mathrm{f}}$, or
\begin{equation}
\Gamma = \frac{1}{\kappa d_\mathrm{f}}.  
\label{eqngamma}
\end{equation}
The prediction for $\Gamma$ relies on the first-passage networks being
fractal.  In Fig.~\ref{fractal} (a), we test Eq.~\ref{fractaldyn} and
observe that $\Delta c$ grows as a power law at large $t$ for all
temperatures studied.  We average $\Delta c$ over $1500$ first-passage
networks and only include $t<\tau_f$ for each realization.  The
exponent $\kappa$ decreases with $T$, which implies that colder systems
explore chemical distance more slowly.

In Fig.~\ref{fractal} (b) we test Eq.~\ref{fractaldim} and find
that, over the limited range of chemical distance accessible to our
small heteropolymer, the chemical fractal dimension $d_f$ is
well-defined and depends linearly on temperature.  $N(\Delta c)$ is
computed by including all sampled basins in $850$ different
first-passage networks at each $T$.  While power-law scaling of $N(\Delta c)$ only
holds for $\Delta c \lesssim 8$, the average chemical distance
explored on a first-passage network is always smaller than $8$.
Therefore, the prediction for $\Gamma$ based on power-law scaling
should hold during the folding process.  In Fig.~\ref{gammalambda}, we
find excellent agreement between the folding-time exponent $\Gamma$
and our prediction $1/\kappa d_\mathrm{f}$.

We have studied first-passage networks formed by the folding
trajectories of a heteropolymer and observed power-law scaling between
the folding time ($\propto N_t$) and number of nodes $N_i$ and bonds
$N_b$ in first-passage networks.  We have also demonstrated that
the folding-time exponent $\Gamma$ can be obtained by measuring the
fractal exponents that characterize the structure of first-passage
networks in configuration space.

Our results do not describe properties of the complete network of
basins in the energy landscape.  However, as far as folding is
concerned, our results suggest that this network is not relevant.
Just as normal diffusion will trace out a two-dimensional fractal
network of sampled nodes, no matter how large the dimension of the
underlying space is, proteins with rugged energy landscapes also trace
out fractal networks that are independent of the complete network.
This behavior is not peculiar to proteins with rugged energy
landscapes, but is also expected in glass-forming materials at low
temperature~\cite{gppglass}.  Moreover, $d_\mathrm{f}$ decreases with
temperature, and is always much smaller than the dimension of
configuration space $D$, which implies that $N_i \sim (\Delta c)^{d_f}
\ll (\Delta c)^D$.  This provides a mechanism by which systems with
rugged energy landscapes can fold reliably without kinetic pathways
and offers a novel resolution to Levinthal's paradox~\cite{levinthal}.

Financial support from NSF grant numbers CBET-0348175 (GL,JB),
DMS-0835742 (CSO), and DMR-0448838 (CSO), and Yale's Institute
for Nanoscience and Quantum Engineering (GL) is acknowledged.


\begin{thebibliography}{100}
\bibitem{footnote2} A contact is defined when the distance between two
monomers is less than the separation at which the concavity of the
Lennard-Jones potential changes sign.
\bibitem{footnote1} We use $\Delta c$ because the Euclidean distance
between basins in configuration space is not uniquely defined.
\end{thebibliography}
\end{document}